\documentclass[pre,preprintnumbers,amsmath,amssymb,twocolumn]{revtex4}
\usepackage{graphicx}% Include figure files
\usepackage{bm,color}% bold math
\usepackage{caption}
\usepackage{mathrsfs}
\usepackage{verbatim} 
\usepackage{tikz}

\definecolor{labelkey}{cmyk}{.4,.2,0,0}

\newcommand{\nn}{\nonumber}

\begin{document}

\title{Relevance of electron spin dissipative processes on Dynamic Nuclear Polarization {\em via} Thermal Mixing}
\author{\bf  Sonia Colombo Serra$^1$, Marta Filibian$^{2}$, Pietro Carretta$^{2}$, Alberto Rosso$^{3}$ and Fabio Tedoldi$^1$}

\affiliation{\medskip
$^{1}$Centro Ricerche Bracco, Bracco Imaging Spa, via Ribes 5, 10010 Colleretto Giacosa (TO), Italy. \\
$^{2}$Dipartimento di Fisica and Unit$\grave{a}$ CNISM, Universit$\grave{a}$ di Pavia, 27100 Pavia, Italy. \\
$^{3}$Universit\'e Paris-Sud, CNRS, LPTMS, UMR 8626, Orsay F-91405, France.\smallskip}

\begin{abstract}
The available theoretical approaches aiming at describing Dynamic Nuclear spin Polarization (DNP) in solutions containing molecules of biomedical interest and paramagnetic centers are not able to model the behaviour observed upon varying the concentration of trityl radicals or the polarization enhancement caused by moderate addition of gadolinium complexes. In this manuscript, we first show experimentally that the nuclear steady state polarization reached in solutions of pyruvic acid with 15 mM trityl radicals  is substantially independent from the average internuclear distance. This evidences a leading role of electron (over nuclear) spin relaxation processes in determining the ultimate performances of DNP. Accordingly, we have devised a variant of the Thermal Mixing model for inhomogenously broadened electron resonance lines which includes a relaxation term describing the exchange of magnetic anisotropy energy of the electron spin system with the lattice. Thanks to this additional term, the dependence of the nuclear polarization on the electron concentration can be properly accounted for. Moreover, the model predicts a strong increase of the final polarization on shortening the electron spin-lattice relaxation time, providing a possible explanation for the effect of gadolinium doping.
\end{abstract}

\maketitle
 
\section{Introduction}

Among the different techniques which allow the nuclear spin polarization to be enhanced to almost its maximum theoretical value, Dynamic Nuclear Polarization (DNP) is raising in popularity. 
The method is flexible enough to be applied to a variety of molecules of biological interest that in recent years has catalyzed dramatic advances for {\em in vivo} $^{13}$C Magnetic Resonance Imaging (see the reviews \cite{BA3, dutta}). DNP increases the nuclear steady state polarization through a transfer of spin order between the electron and the nuclear spin systems, occurring when the Electron Spin Resonance (ESR) line is suitably irradiated. For biomedical imaging purposes this transfer takes place among the electrons of stable radicals and the nuclei of biomolecules, in a solution which is cooled to $T\simeq 1$ K. Once the nuclear polarization process has taken place, the frozen solution is rapidly dissolved (while maintaining most of the spin order just created \cite{PNASJHAL}) and injected into living subjects to eventually image  the metabolic fate of the hyperpolarized substrates {\em in vivo}.

In parallel to the development of these novel biomedical applications, a renewed commitment towards the understanding of the physical mechanisms driving DNP is emerging. The basic physical concepts underlying DNP phenomenology, have already been described few decades ago (see \cite{AbragamGoldman} and reference therein) and three different regimes, the Solid
Effect, the Cross Effect and the Thermal Mixing (TM) regime, were specified according to the typical parameters of the system, such as the nuclear resonance frequency, the strength of the interaction between the spins and the magnitude of the external magnetic field. Considerable steps forward have recently been made in the quantum mechanical description of Solid Effect \cite{Vega1, Vega2, kock} and Cross Effect \cite{Vega3, Vega4}. The relevant regime for biomedical application, however, has been argued to be the TM regime \cite{JHAL2008}, with a dipolar interaction among the electron spins which is stronger than their coupling with the lattice, although very small with respect to the $g$-anisotropy terms responsible for the broadening of the electron resonance line.

The traditional approach to the TM regime is based on an effective thermodynamic model (the so-called spin temperature approach): such description, while providing a qualitative picture of the expected steady state polarization under microwave saturation, does not include any dependence on the strength of the interactions between nuclear and electron spins, as well as on the relative intensities between those interactions, the spin-lattice relaxation rates and the microwave irradiation power. Improvements to the original theoretical picture have been proposed in \cite{JanninMW} and \cite{TM1, TM2}. In these latter papers, in particular, a novel approach based on rate equations has been introduced, that overcomes several limitations of the traditional approaches and provides the dynamics of the spin polarization of spin ensembles by calculating the evolution induced by any single or multiple spin transitions. 
There are however certain experimental observations (see Appendix \ref{review} for details) pointing out a complexity of the DNP phenomenon that remains largely unexplained. The typical DNP formulations that guarantee an adequate polarization level for {\em in vivo} procedures are normally obtained using relatively low concentration of trityl radicals (between 10 and 20 mM), but no exhaustive explanation of why a higher concentration of electron spins causes a reduction of the final nuclear polarization was given so far. Moreover, according the observation first reported in \cite{Thanning, JHAL2008}, trace amounts (1-2 mM) of gadolinium complexes added to the solutions can further improve the DNP signal enhancement. The addition of gadolinium, now commonly exploited in standard protocols for DNP sample preparation \cite{citA, citB, citC}, was shown to affect neither the electron linewidth nor the nuclear spin-lattice relaxation time, while it induces a significant reduction of the electron spin relaxation time $T_{1e}$ \cite{JHAL2008, JHAL2010, LumataEL}. In the Borghini's framework, however, such reduction would affect the nuclear polarization by less than 10 \% \cite{JHAL2008}, whereas the gadolinium-induced enhancement observed experimentally is up to four-fold.

In this paper we first integrate the available experimental scenario with new data, collected on a prototype sample (Section \ref{nconc}), showing how the final value of the nuclear polarization does not depend on the concentration of nuclear spins. Inspired by such observation, in Section \ref{TheoreticalModel} we introduce a variant of the rate equation approach proposed in \cite{TM1}, which includes a dissipative term within the electron system controlled by the radical concentration. The numerical predictions of this novel model, reported in Section \ref{NumericalResults}, reproduce rather well the experimental behaviour of the nuclear polarization {\em versus} electron spin density as well as its extreme sensitivity to  the reduction of $T_{1e}$. These and other aspects of the comparison between theoretical and experimental results (these latter recalled in Appendix \ref{review}) are discussed in more detail in Section \ref{conclusion}. All technical details of the model computation have been included in Appendix \ref{appA} and \ref{appC}, to better highlight the main messages of this work.

\section{Dependence of steady state polarization on nuclear concentration}
\label{nconc}

\begin{figure}[b]
 \includegraphics[width=7.6 cm]{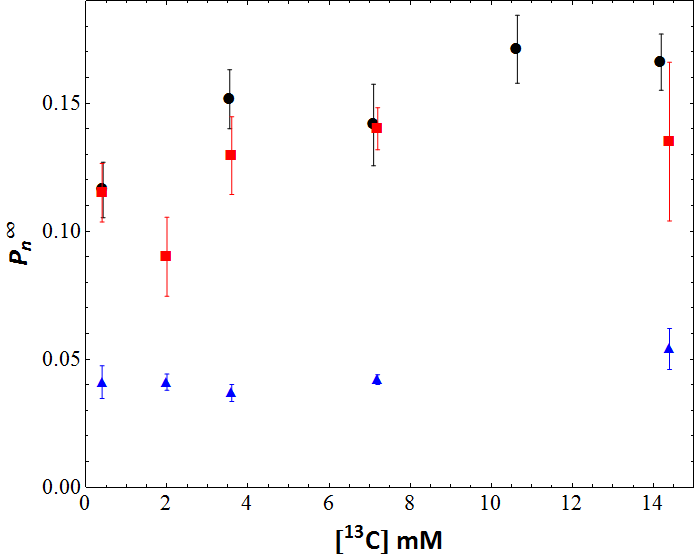}
\caption{Optimal $P_n^{\infty}$ as a function of the $^{13}$C concentration for a sample of pyruvic acid doped with OX063 trityl radical 15 mM measured at $B = 3.35$ T / $T = 1.2$ K (circles), at $B = 3.46$ T / $T = 1.8$ K (squares) and $B = 3.46$ T / $T = 4.2$ K (triangles).}
\label{nuclei}
\end{figure}

Traditional models assuming a perfect thermal contact between electron and nuclear spins and an energy exchange between the spin systems and the lattice occurring via $T_{1e}$ and $T_{1n}$ only, overestimate the steady state nuclear polarization $P_n^{\infty}= P_n(t \rightarrow \infty)$. Thus, an improved TM theory should rely onto different dissipative scenario. In order to clarify whether these mechanisms involve primarily the electron or the nuclear spin reservoir, we have experimentally investigated the modifications of $P_n^{\infty}$ in pyruvic acid samples doped with OX063 trityl radical $15$ mM at variable $^{13}$C concentration.

To keep all the other properties of the solution unchanged, samples with different $^{13}$C concentrations were obtained by mixing unlabeled pyruvic acid and [1-$^{13}$C]pyruvic acid in different ratios (100 \% labeled sample, corresponding to 14.2 M $^{13}$C concentration;  75\% labeled,  10.6 M;  50 \%,  7.1 M;  25\% labeled,  3.55 M;  10 \% labeled, 1.42 M;  unlabeled sample, corresponding to 0.43 M $^{13}$C concentration). The carbon nuclear system of the unlabelled product is made by $\approx 99 \%$ spinless $^{12}$C, whereas in fully labelled [1-$^{13}$C]pyruvic acid, 1/3 of the carbon nuclei have spin $S$ = 1/2, since each pyruvic molecule has 3 carbons but only those in position-1 are $^{13}$C enriched.%2.5

$^{13}$C polarization measurements were performed using two different apparati operating at about the same magnetic field $B$ but with different capabilities of temperature-regulation, to check whether or not the information we look for is temperature dependent. The first DNP system operates at $B = 3.35$ T / $T = 1.2$ K and is equipped with a $0$-$200$ mW microwave (MW) source that can be sweeped between $93.75$ and $94.25$ GHz and with a $35.86$ MHz radiofrequency (RF) set up. The second one works at variable temperature in the $1.8$-$4.2$ K range and uses a 32 mW Gunn Diode MW Source working in the range $95.96$-$98.04$ GHz and a $37.05$ MHz RF probe. An amount of about 100 mg of each sample underwent flash freezing in a cryogenic bath before starting MW irradiation operating at the frequency corresponding at the maximum enhancement. The $^{13}$C NMR signal build up was sampled after RF saturation (in order to destroy any residual signal) up to steady state by means of low flip angle ($\alpha$ about 6 $^{\circ}$) acquisitions \cite{PNASJHAL}. $P_n^{\infty}$ and the polarization time constant $T_{\text{pol}}$ were derived by fitting the build up curves to an expression that takes into account the reduction of the $^{13}$C signal amplitude with time induced by the readout pulses according to ref. \cite{RFcorrection}.

For low $^{13}$C concentrations the evolution of the nuclear polarization turned out to be very slow (at $1.2$ K $T_{\text{pol}} \approx 4000$ s for the sample at natural abundance {\em vs} $\approx 1000$ s for the fully labelled sample), reflecting a degradation of the electron-nucleus contact on increasing the average distance between the two spin species and a slow nuclear spin diffusion. However, at all the investigated temperatures the final value of $P_n^{\infty}$ was found to be nearly independent from the $^{13}$C concentration [$^{13}$C]  (\figurename~\ref{nuclei}), similarly to what was reported also for other samples polarized by means of nitroxides radicals \cite{Comment}. This suggests that $P_n^{\infty}$ is substantially unaffected by nuclear relaxation mechanisms that do not involve the electron reservoir. In fact, as remarked in \cite{TM2}, in presence of a sizeable direct exchange between the nuclear system and the lattice, one would expect a substantial reduction of $P_n^{\infty}$ on decreasing the electron-nucleus contact. Thus, it is concluded that the reduction of $P_n^{\infty}$ should primarly originate from dissipative processes involving the electron spins.

\section{Theoretical Model} \label{TheoreticalModel}

\begin{figure}[t]
\includegraphics[width=9 cm]{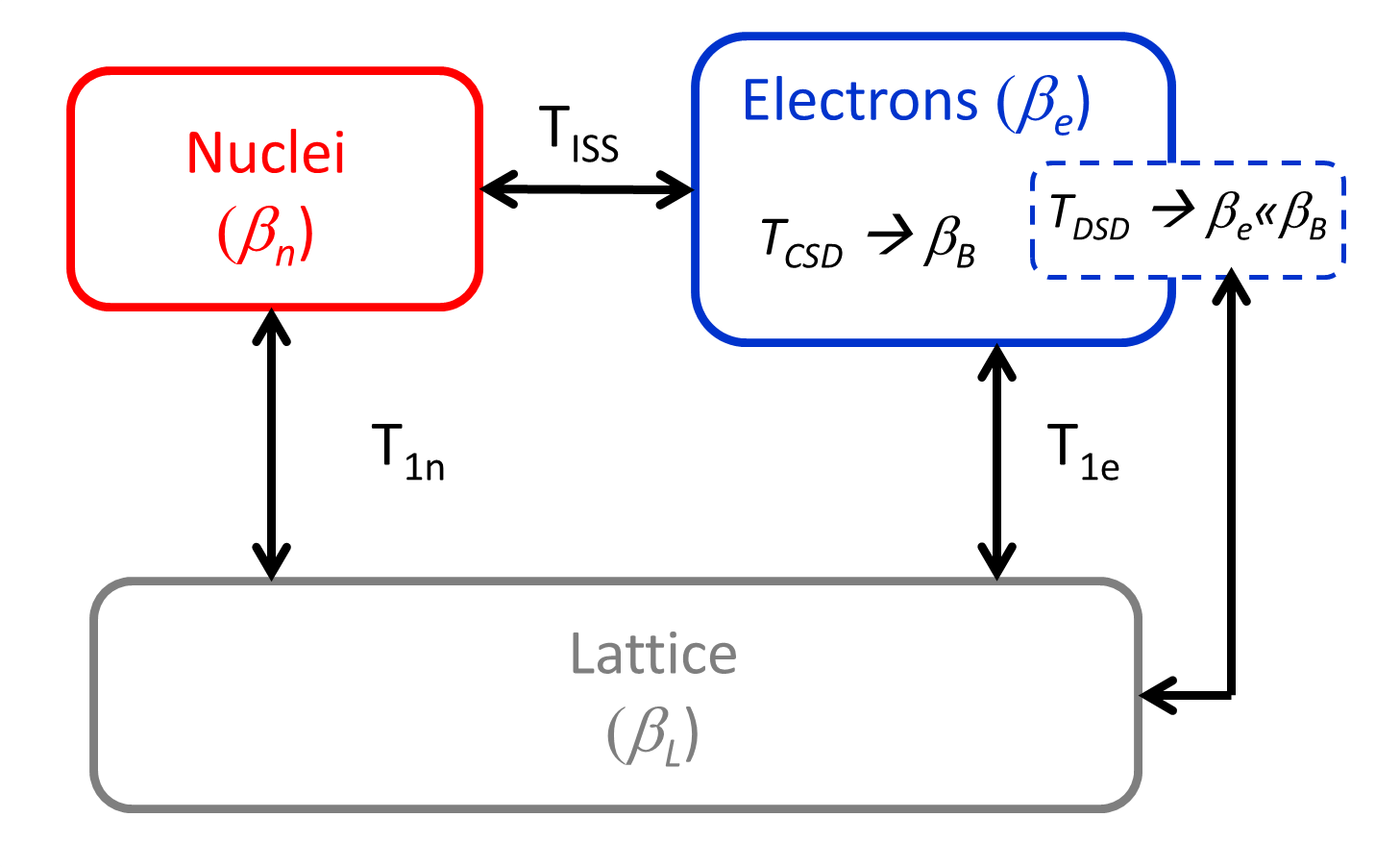}
\caption{Thermal systems and interactions involved in TM-DNP. Nuclei directly feel the lattice through the leakage term $T_{\text{1n}}$ which represents the nuclear relaxation processes not mediated by electrons. {\em Via} the three particle mechanism $T_{\text{ISS}}$ electrons are in contact with the nuclear system, while thermalizing internally by energy conserving spectral diffusion ($T_{\text{CSD}}$) and with the lattice by Zeeman transitions ($T_{\text{1e}}$). Moreover, in the model presented here, electron spins also interact among themselves and with the lattice through dissipative spectral diffusion ($T_{\text{DSD}}$).}
\label{FIGURESystem}
\end{figure}

\begin{figure}[t]
 \includegraphics[width=8 cm]{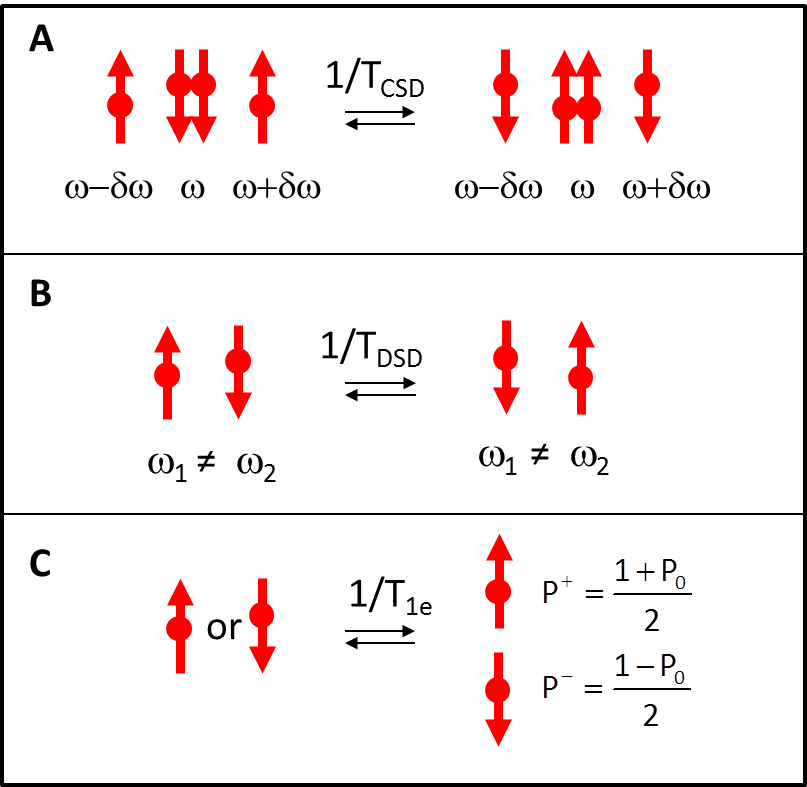}
\caption{Microscopic interactions driving the evolution of the electron polarization profile under the assumption of bad contact between electrons and nuclei and negligible leakage ($T_{\text{ISS}}$, $T_{1n}$ $ \rightarrow \infty$).
 A: energy conserving spectral diffusion. B: dissipative spectral diffusion (electron electron flip-flop). C: electron spin-lattice relaxation towards the Boltzaman equilibrium polarization $P_0 = \tanh\left[\beta_L \omega_e\right]$.}
\label{FIGURETransition}
\end{figure}

The TM regime is characterized by a large spread of electron Larmor frequencies (larger than the nuclear Larmor frequency $\omega_n$), arising from a distribution of local magnetic fields as, for example, in presence of a $g$-tensor anisotropy. In these conditions it is useful to split the electron population into spin packets sharing the same Larmor frequency. The sum of these packets yields the ESR line. At a given frequency $\omega$ the intensity of the ESR line $f(\omega)$ is proportional to the number of electrons resonating at that frequency. The function $f(\omega)$ and the average electron frequency  $\omega_e$ are defined so that $\int f(\omega) d\omega = 1$ and $\int (\omega-\omega_e) f(\omega) d\omega = 0$. %2.12
At equilibrium condition the electron system assumes a nearly constant polarization profile: 
\begin{equation}
P_e^0(\omega) \approx P_e^0(\omega_e) = P_0 = \tanh \left[\beta_L \omega_e \right],
\label{eqpw}
\end{equation}
where $\beta_L = \hbar/\left(2 k_B T_L\right)$ is the inverse lattice temperature (notice that, as here, it is normally reported in time units). In a magnetic field strength of the order of 1 T at a temperature $T_L$ of 1 K, $P_0 \approx 1$.

Under microwave irradiation the electron spin system moves from Boltzman equilibrium to a non-equilibrium non-uniform steady state, characterized by a frequency-dependent profile $P_e^{\infty}(\omega)= \tanh \left[\beta_e \omega_e \right]$ (with $\beta_e \gg \beta_L$ according to the spin temperature approach), which is responsible for the enhancement of the nuclear polarization $P_n^{\infty} = \tanh \left(\beta_n \omega_n\right)$. A pictorial description of this phenomenon and on the three different subsystems involved is sketched \figurename~\ref{FIGURESystem}.

In general, the computation of the steady state electron inverse temperature $\beta_e$ is a rather diffcult task since the electron spins, which are strongly interacting with each other {\em via} dipolar coupling, are out of equilibrium due to MW irradiation and are at the same time in contact with a thermal bath (lattice). Under certain assumptions, discussed in detail below, Borghini \cite{AbragamGoldman, BorghiniPRL} was able to compute $P_e^{\infty}(\omega)$ and, accordingly,  its corresponding inverse temperature $\beta_e = \beta_B$, and to provide an upper bound for $P_n^{\infty}$. His overestimation is particularly evident when MW irradiation is performed at the edges of the ESR line where one would expect to observe an almost negligible polarization enhancement. %2.14
In previous works \cite{TM1, TM2} we showed that a finite electron nucleus contact, even in the mean field approximation, allows to recover realistic values of the nuclear spin polarization. In that model the finite electron-nucleus contact combines both the ISS processes (a simultaneous flip-flop of two electron spins compensated by a nuclear spin flip) and the nuclear spin diffusion. Nuclei reach an intermediate inverse  spin temperature between $\beta_L$ and $\beta_B$, corresponding to a reduced $P_n^{\infty}$, and determined by the ratio between the electron-nucleus contact (quantified by $T_{\text{ISS}}$) and the nuclear leakage ($T_{1n}$). By properly tuning the values of these parameters, one reproduces MW spectra similar to those observed in experiments. However, according to such a model one would expect $T_{\text{ISS}}$ to increase both with the nuclear and with the electron spin concentration that, by improving the contact among the two spin systems, would reduce the relative efficiency of the leakage, leading to higher steady state polarization levels. Since this does not correspond to the behaviour observed experimentally (see also Section \ref{nconc}), it becomes necessary to go beyond the profile $P_e^{\infty}(\omega)$ proposed by Borghini and search for different electron steady states, characterized by an inverse temperature $\beta_e$ smaller than $\beta_B$ and dependent on the electron spin density.

First of all, it has to be realized that the Borghini model and the model introduced in \cite{TM1, TM2} rely on the assumption that the energy exchange between the electron system and the lattice occurs only {\em via} the Zeeman transitions depicted in \figurename~\ref{FIGURETransition}, panel C, whereas all transitions involving more than one electron spin are always
energy conserving. The simplest microscopic process of this kind, characterized by a time-scale $T_{\text{CSD}}$ \footnote{Conversely to what we did in the past \cite{TM1, TM2}, we chose here to use a different symbol for the typical time of this four electron mechanism previously named $T_{2e}$. This in order to better distinguish the microscopic events ($T_{\text{CSD}}$) partially responsible for the energy exchange between different electron spins from the collective time (usually reported as $T_{2e}$) that describes the spin-spin relaxation from a thermodynamic point of view.}, is depicted in \figurename~\ref{FIGURETransition}, panel A. Transitions involving more than one spin are the elementary events of the phenomenon referred to as spectral diffusion in the low temperature TM-DNP description proposed in \cite{AbragamGoldman} and here named energy conserving spectral diffusion (CSD).
When CSD is infinitely efficient ($T_{\text{CSD}} = 0$), the electron system is driven towards a high inverse temperature $\beta_B$.

In this manuscript we consider an alternative model based on three main realistic assumptions. {\em First}, the energy conserving spectral diffusion is not, as it was always assumed so far in the TM models, infinitely fast ($T_{\text{CSD}} \neq 0$). {\em Second}, non conserving electron flip-flop processes (characterized by a time constant $T_{\text{DSD}}$) are possible. The energy required for the electron flip flop in a DSD event must be provided or adsorbed by the lattice thermal bath. The most elementary non conserving transition is represented in \figurename~\ref{FIGURETransition}, panel B. These latter events, analogously to $T_{\text{CSD}}$ processes, promote an internal thermalization among the different spin packets of the ESR line and for this reason we refer to their macroscopic effect as {\em dissipative} spectral diffusion (DSD). Since at T approx. 1 K the lattice is lacking in its capabilities of emitting energy quanta (even if at very low frequency) that can stimulate spin transitions, we expect $T_{\text{DSD}}$ to be much longer than $T_{\text{CSD}}$, the latter not requiring any energy exchange of the electron spin system  with the lattice.
The {\em third} assumption originates from the experimental results reported in the previous section, which clearly show that when trityls are exploited as polarizing agents for $^{13}$C nuclei, the electron-nucleus contact $1/T_{\text{ISS}}$ and the nuclear leakage $1/T_{1n}$ are weak enough to make any electron polarization loss {\em via} the nuclear channel irrelevant in defining the final steady state of electrons. In this limit $P_e^{\infty}(\omega)$ is determined only by the competition between energy conserving and non conserving processes and can be derived, upon considering the electron spins as a fully connected system, by solving a system of mean field rate equations (Appendix \ref{appA}) describing the three processes in \figurename~\ref{FIGURETransition}. Once $P_e^{\infty}(\omega)$ is known, the corresponding nuclear polarization $P_n^{\infty}$ is derived as described in \cite{TM2} (see Eq.(9)) through the following expression:

\begin{equation}
P_n^{\infty} = \frac{\int f(\omega) f(\omega + \omega_n) \left[P_e^{\infty}(\omega) - P_e^{\infty}(\omega + \omega_n) \right] d \omega}{\int f(\omega) f(\omega + \omega_n) \left[ 1- P_e^{\infty}(\omega) P_e^{\infty}(\omega + \omega_n)\right] d\omega} .
\label{eqpn*}
\end{equation}

It is worth to remark that  the first  assumption of the model  ({\em i.e.} finite $T_{\text{CSD}}$) defines a regime which is not a pure TM ({\em i.e.} the electron profile is in general not a simple hyperbolic tangent function with a defined spin temperature $\beta$). In the numerical calculations that follows however, the interactions between electrons are generally assumed to be strong enough to impose  an electron profile which is significantly different form the equilibrium condition. By increasing $T_{\text{CSD}}$ and $T_{\text{DSD}}$ one may explore different regimes and in particular, for $T_{\text{CSD}}$, $T_{\text{CSD}} \rightarrow \infty$ and $T_{\text{ISS}} \ll \infty$, a Cross Effect behavior is recalled.

\section{Numerical Results}
\label{NumericalResults}

\begin{figure}[t]
 \includegraphics[width=8 cm]{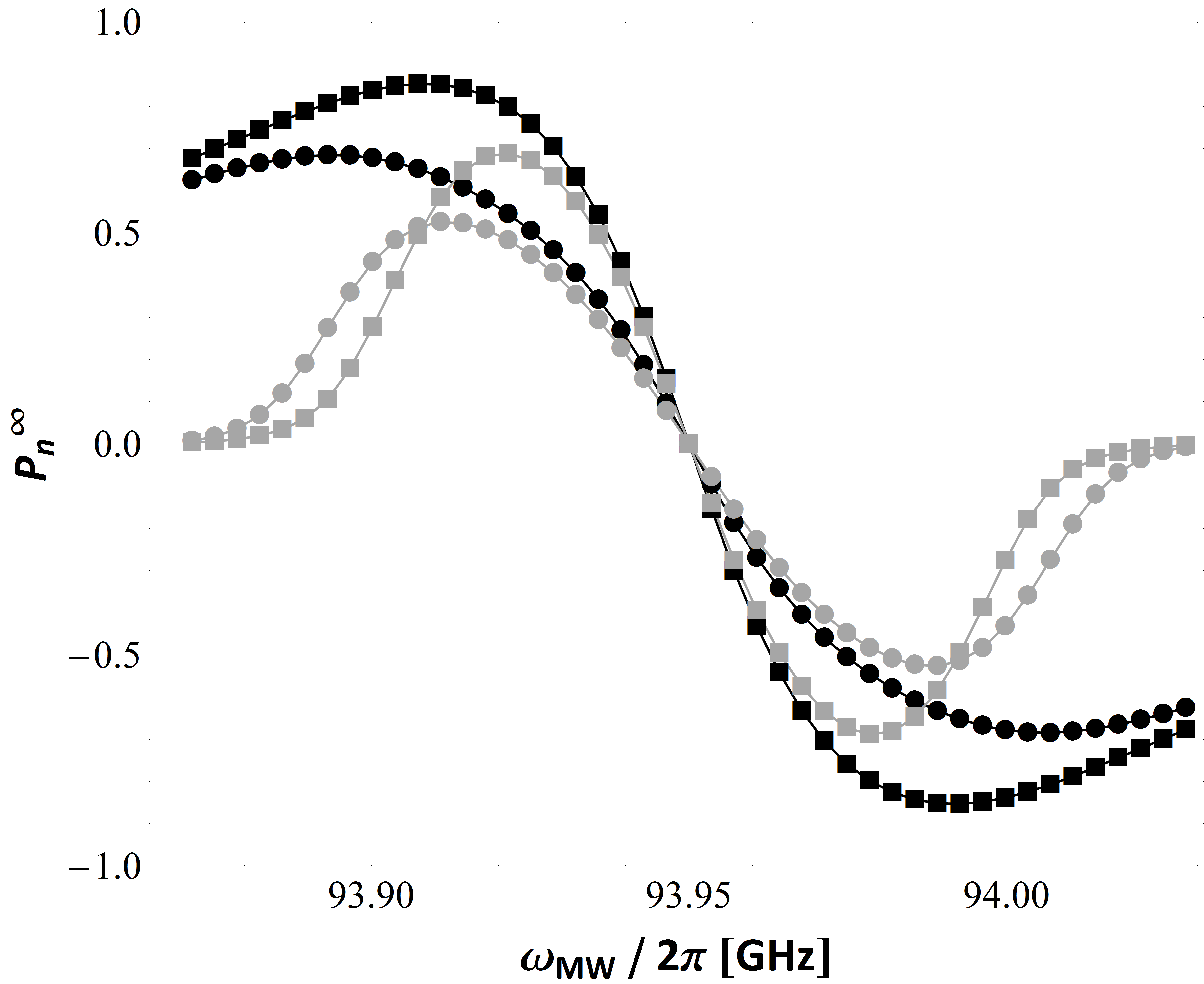}
\caption{ $P_n^{\infty}$ MW spectra on varying $T_{\text{CSD}}$ and  $T_{\text{DSD}}$. Black squares: Borghini model. Grey squares: behaviour obtained from  Eq.s \ref{eqpn*} and \ref{rateeqT2e} with $T_{1e}=1$ s, $T_{\text{CSD}}=10^{-7}$ s and  $1/T_{\text{DSD}}=0$. Black circles: from Eq.s \ref{eqpn*} and \ref{rateeqT2e} with $T_{1e}=1$ s, $T_{\text{CSD}}=0$ s and $T_{\text{DSD}}=10^{-3}$ s. Grey circles: from Eq.s \ref{eqpn*} and \ref{rateeqT2e} with $T_{1e}=1$ s, $T_{\text{CSD}}=10^{-7}$ s and  $T_{\text{DSD}}=10^{-3}$ s.}
\label{Figure2NR}
\end{figure}

\begin{figure}[t]
 \includegraphics[width=8 cm]{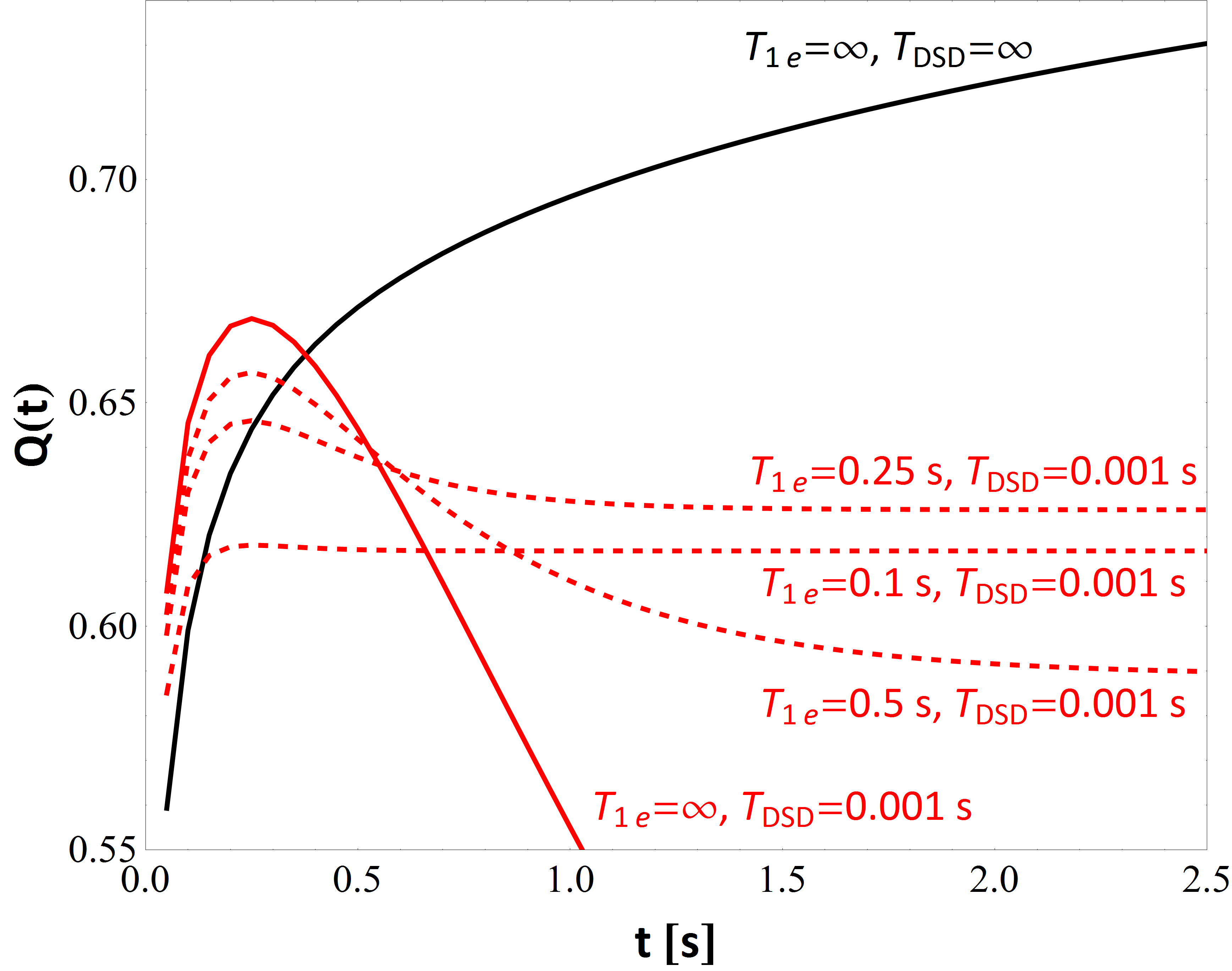}
\caption{Time evolution of the electron profile encoded in the scalar variable $Q(t)$ computed according to Eq.(\ref{eqQ}) upon saturating the most effective packet and setting $T_{\text{CSD}} = 10^{-7}$ s.}
\label{Figure1NR}
\end{figure}

Electron and nuclear polarizations have been computed by numerically solving the rate equation system (\ref{rateeqT2e})  and Eq. (\ref{eqpn*}) respectively.
The external magnetic field was set to $B_0 = 3.35$ T and the temperature to $1.2$ K, in agreement with most of the experiments described in literature. To model the ESR line of trityl radicals, a Gaussian distribution centered at $\omega_0 /2 \pi = 93.95$ GHz and with a linewidth of $64$ MHz \cite{JHAL2010} has been used.

Let us first look at the effect of a finite  $1/T_{\text{CSD}}$ rate and of a non vanishing dissipative spectral diffusion $1/T_{\text{DSD}}$ on the steady state nuclear polarization. In \figurename~\ref{Figure2NR},  $P_n^{\infty}$ as a function of the MW irradiation frequency (MW spectrum) obtained under Borghini's assumptions in absence of nuclear leakage ($ T_{\text{CSD}} = T_{\text{ISS}} = 1/T_{1n} = 1/T_{\text{DSD}}=0$, black squares) is plotted together with the output of our computation for the following choices of parameters:

\begin{itemize}
\item $T_{1e}=1$ s, $T_{\text{CSD}}=10^{-7}$ s and  $1/T_{\text{DSD}}=0$, grey squares;
\item  $T_{1e}=1$ s, $T_{\text{CSD}}=0$ s and  $T_{\text{DSD}}=10^{-3}$ s, black circles;
\item  $T_{1e}=1$ s, $T_{\text{CSD}}=10^{-7}$ s and  $T_{\text{DSD}}=10^{-3}$ s, grey circles.
\end{itemize}

One immediately recognizes that the assumption of a non zero, although very small, value of $ T_{\text{CSD}}$ leads to a clipping of the wings of the nuclear polarization spectrum and thus to the overtaking of the most evident limitation of the Borghini model that was encountered also in the high temperature TM limit first discussed by Provotorov \cite{Provotorov1, Provotorov2, Provotorov3, Provotorov4}. A similar improvement is achievable also by assuming a combination of bad electron-nucleus contact and finite nuclear leakage  \cite{TM2} or by imposing a partial saturation of the irradiated packet  \cite{JanninMW, TM2}. Here however it is  obtained simply by removing the non-physical assumption  $T_{\text{CSD}} = 0$, which is normally accepted in all TM models in order to simplify their computation. 
As expected, the second new element of the proposed model (electron electron flip-flop), acts as a source of dissipation within the electron system, producing an overall suppression of the final nuclear polarization. Also in this case a $T_{\text{CSD}} \neq 0$ is mandatory to obtain a sharper and realistic shape of the $P_n^{\infty}$ spectrum.

In order to analyze the time evolution of the electron profile $P_e(\omega, t)$, as obtained by the system of rate equations (\ref{rateeqT2e}), it is convenient to introduce a scalar parameter $Q(t)$ computed from $P_e(\omega, t)$ through the following equation:

\begin{equation}
Q(t) = \frac{\int f(\omega) f(\omega + \omega_n) \left[ P_e(\omega, t) - P_e(\omega + \omega_n, t) \right] d \omega}{\int f(\omega) f(\omega + \omega_n) \left[ 1- P_e(\omega, t) P_e(\omega + \omega_n, t)\right] d\omega} .
\label{eqQ}
\end{equation}

As far as the electron-lattice coupling ($1/T_{1e}$) is widely more efficient than the electron-nucleus contact ($1/T_{\text{ISS}}$), $Q(t)$ represents the nuclear polarization that would be associated to the electron profile $P_e(\omega, t)$ if this latter  was a steady state profile.%2.19
It is important to realize that $Q(t)$ and $P_n(t)$ have rather different build up times, being the latter dependent on $T_{\text{ISS}}$, and converge to the same value only for $t\rightarrow \infty$.

When $T_{\text{DSD}} \rightarrow \infty$ the electron profile under MW irradiation at frequency $\omega_{MW}$ progresses from hole burning (where the irradiated packet is saturated and all other packets are at Boltzman equilibrium) to a profile which approaches, in the limit $T_{\text{CSD}} \ll T_{1e}$, the one predicted by Borghini ($P_e(\omega) = \tanh \left[\beta_B (\omega-\omega_{MW})\right]$). In  the example depicted in \figurename~\ref{Figure1NR} in particular, the system is shown to evolve from $Q(t=0) = 0.34$ to $Q(t\rightarrow\infty) = 0.825$. When $T_{\text{DSD}}$ is finite and the electron spin-lattice relaxation is negligible ($T_{1e} \rightarrow \infty$), the system starts to evolve from the hole burning profile ($Q(t=0) = 0.34$) to larger $Q$ values up to $t \approx T_{\text{DSD}}$, when the dissipative process becomes relevant and the saturation of the burned spin packet slowly spreads throughout the ESR spectrum ($Q(t=\infty) = 0$). This catastrophic fate is prevented by the onset of electron spin lattice relaxation which freezes the electron profile for $t > T_{1e}$. In summary two time-regimes can be identified:
\begin{itemize}
\item $t \ll T_{\text{DSD}}$, where CSD processes dominate and increase the nuclear polarization;
\item $t \gg T_{\text{DSD}}$, where DSD processes are effective and reduce $P_n^{\infty}$.
\end{itemize}
Both the reduction of $P_n^{\infty}$ observed when increasing the radical concentration and the polarization enhancement following gadolinium doping find a natural explanation within this general framework.
By shortening $T_{\text{DSD}}$ and $T_{\text{CSD}}$ without significantly affecting $T_{1e}$, a large number of paramagnetic centers decreases the nuclear polarization. On the other hand, any perturbation that solely reduces $T_{1e}$ leaving the spectral diffusion parameters unchanged (as gadolinimu doping is expected to do), has a positive outcome on  $P_n^{\infty}$.

\begin{figure*}[htbp]
 \includegraphics[width=17.6 cm]{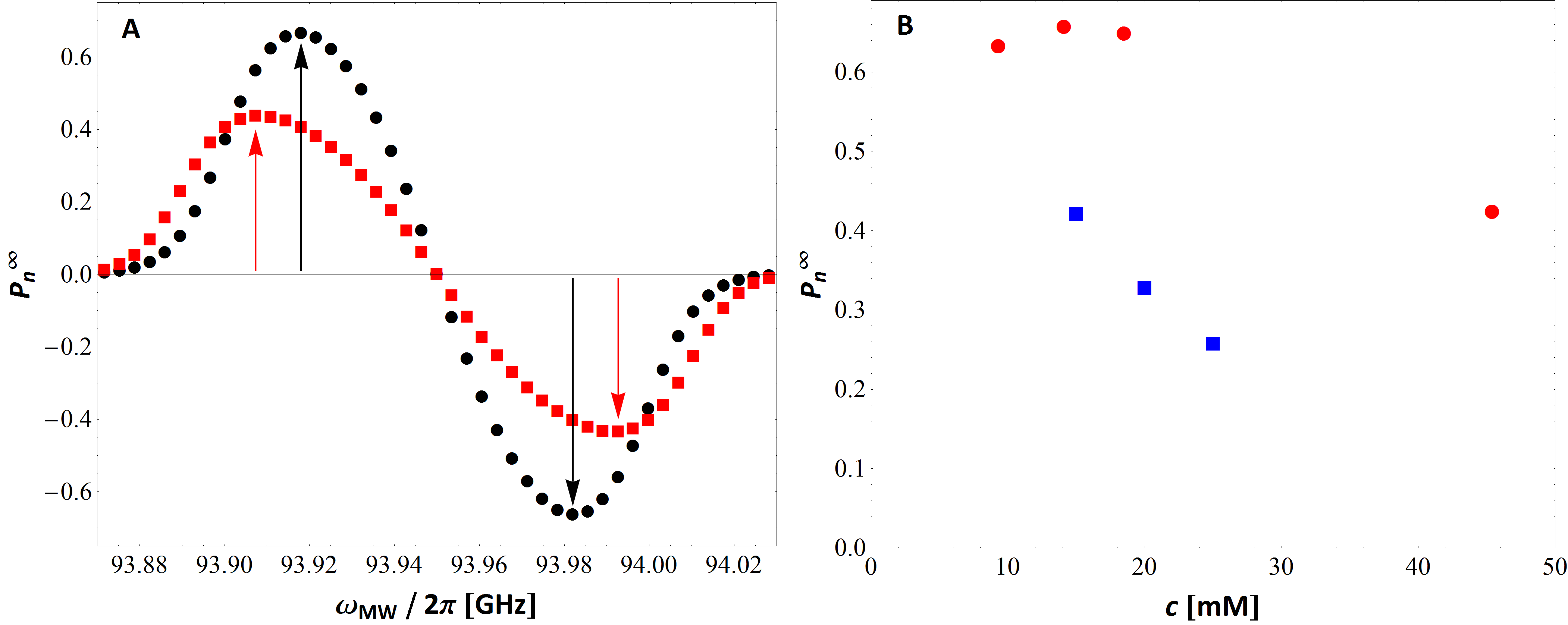}
\caption{Effect of electron concentration. Simulated data for systems with $T_{1e} = 1$ s. Panel A, circles:  MW spectrum obtained by setting $T_{\text{CSD}}$ = $10^{-7}$ s and $T_{\text{DSD}} $ = $5~10^{-3}$ s, , to reproduce the behaviour of a [1-$^{13}$C]pyruvic acid sample with trityl $15$ mM \cite{JHAL2008, JHAL2010}. Panel A, squares:  MW spectrum obtained by scaling  $T_{\text{CSD}}$ and $T_{\text{DSD}} $ with the electron spin concentration as specified in the text, to predict the behaviour of a [1-$^{13}$C]pyruvic acid sample with trityl $45$ mM. Panel B: Maximum $P_n^{\infty}$ as a function of the electron spin concentration $c$ obtained by scaling as described in the text the $T_{\text{CSD}}$ and $T_{\text{DSD}} $ values fitted to the  [1-$^{13}$C]pyruvic acid sample with trityl $15$ mM (circles). The squares represent the behaviour obtained by setting $T_{\text{CSD}}$ = $10^{-7}$ s, $T_{\text{DSD}}$ = $5~10^{-4}$ s to match the experimental polarization of a [$^{13}$C]urea sample with trityl $15$ mM, and by scaling them with concentration similarly to the case of [1-$^{13}$C]pyruvic acid. The simulated trends reflect the corresponding experimental behaviours reported in the insets of \figurename~\ref{figureAp}.}
\label{figureApsimul}
\end{figure*}

\subsection{Effect of electron concentration}

An increase of the electron concentration $c$ is expected to lead to an enhancement of the transition rates $1/T_{\text{CSD}}$ and $1/T_{\text{DSD}}$ that depend on the mutual distances between electrons. In our simulations the two parameters were phenomenologically assumed to scale with $c^2$. Beside allowing to nicely reproducing the observed experimental behavior, this choice for $1/T_{\text{DSD}}$ was inspired by the $c^2$ dependence reported by Abragam and Goldman in reference \cite{AbragamGoldman}, Eq. (6.50), for the $ 1/T_{\text{ISS}}$ process. The two mechanisms are in fact analogous from the electron point of view (both involving a flip flop event between two electrons), being only differentiated by the partner for energy exchange: the nuclear system for ISS and the lattice for DSD. The choice of the same scaling function for $1/T_{\text{DSD}}$ does not have instead any  {\em a priori} justification, but it has been checked to not significantly affect the $c$ dependence of $P_n^{\infty}$ as far as $T_{\text{CSD}} < T_{\text{DSD}}$.

Numerical results, obtained by setting $T_{1e} = 1$ s according to \cite{JHAL2008, JHAL2010} and by adapting $T_{\text{CSD}}$ and $T_{\text{DSD}}$ to fit the experimental value $P_n^{\infty}$ of given samples at $15$ mM trityl concentration, are shown in \figurename~\ref{figureApsimul} to properly model the experimental behaviour found in [1-$^{13}$C]pyruvic acid and [$^{13}$C]urea samples described in Appendix \ref{review}, \figurename~\ref{figureAp}. The decrease of $P_n^{\infty}$ observed at high radical concentration for both systems (\figurename~\ref{figureApsimul}, panel B) is due to the higher efficiency of DSD processes in the regime where they are predominant. Conversely, the small increase observed for the [1-$^{13}$C]pyruvic acid sample at low radical concentration corresponds to the regime where CSD dominates.
At very low concentration the polarization shows the correct qualitative behavior but  its absolute value is slightly higher than what was found in experiments. A possible explanation is that our approach assumes, for the limit of vanishing electron concentration, an hole burning shape which corresponds to a pretty high value of $P_n^{\infty}$. This assumption is quite crude because the hole burning shape is not realistic even when the electron packets are not interacting.

Finally, when the whole MW spectra at two different radical concentration are considered (\figurename~\ref{figureApsimul}, panel A), a shift of the maximum enhancement position towards the edge of the spectrum, when increasing {\em c}, is observed (peak to peak distance = 65 MHz for the 15 mM sample and 90 MHz for the 45 mM sample). Since to our knowledge no experimental evidence of such behavior is reported in literature, we  performed a dedicated experiment using the 3.35 T set up described in Section \ref{nconc} to measure the MW spectrum of the 45 mM trityl doped sample \footnote{The sweep was performed from high to low frequency with a delay of 540 s (approx. 3 times the polarization time constant at the optimal frequency) between frequency changes. The signal was recorded every minute by means of low flip angle  (approx.  4$^{\circ}$) acquisitions to monitor the nuclear polarization build-up. The area of the spectrum acquired after  540 s of irradiation at  a given frequency was then plotted as a function of the MW frequency and a peak to peak separation of approx. 100 MHz was extracted.}. A peak to peak distance of approx. 100 MHz was found, that is higher than the one observed in the 15 mM sample, 62 MHz  according to \cite{JHAL2008, JHAL2010}, and qualitatively confirms our theoretical prediction.

\subsection{Effect of gadolinium doping}

The effect of doping the DNP samples with moderate quantities of gadolinium complexes has been modeled by considering the $T_{1e}$ reduction that, according to literature data  \cite{JHAL2010}, follows such doping. For numerical simulations a high (although finite) CSD rate has been set ($T_{\text{CSD}} = 10^{-7}$ s), whereas the spin lattice relaxation was taken from literature \cite{JHAL2008, JHAL2010} ($T_{1e} = 1$ s in absence of gadolinium). $T_{\text{DSD}}$ was adapted to suitable reproducing  $P_n^{\infty}$ MW spectra without gadolinium. The two major effects of reducing $T_{1e}$ are shown in  \figurename~\ref{figureGdsimul}, panel A:  ({\em i}) the overall nuclear polarization is enhanced; ({\em ii}) the peaks position is shifted towards the centre of the spectrum. The quantitative agreement between the simulated MW spectra and their experimental counterpart reproduced in \figurename~\ref{figureGd} (in particular the 4-fold enhancement) is remarkable.
The behaviour of the maximum $P_n^{\infty}$ as a function of $1/{T_{1e}}$ is represented in panel B. 

Two regimes can be recognized: for relatively small reduction of $T_{1e}$ (low gadolinium concentration) the nuclear polarization is enhanced since such a reduction leads to a suppression of the dissipative effect induced by $T_{\text{DSD}}$ processes; for more significant $T_{1e}$ reductions (higher gadolinium concentration) $P_n^{\infty}$ decreases and slowly reaches the value correspondent to the hole burning profile. No assumption was made regarding  the specific dependence of $T_{1e}$ on gadolinium concentration, since  the available experimental data do not allow to establish a clear functional relation between the two parameters.
It can not be excluded \emph{a priori} that the presence of gadolinium affects also other interaction parameters, such as $T_{\text{DSD}}$, and not only $T_{1e}$. The only strong experimental observation available on the effect of gadolinium doping, however, is a significant reduction of $T_{1e}$, that we proved here to be sufficient to account for the correspondent polarization enhancement as well as for the modification of the microwave DNP spectrum.

\section{Discussion and conclusion}
\label{conclusion}

The aim of this work was to find a suitable theoretical justification for the effects of radical concentration and gadolinium doping on the ultimate performances of a DNP procedure carried out a $T \approx 1$ K with trityl radicals. 
The available experimental data show that $P_n^{\infty}$ decreases at high trityl concentration and increases after addition of moderate amounts of gadolinium complexes. The $P_n^{\infty}$ reduction on increasing the number of paramagnetic centers is associated with a faster polarization build up. While a speed up of the DNP process with the increase in the radical concentration has to be expected, since there are more polarization transfer centers, the significant decrease of the steady state polarization has not been accounted for by any previous theoretical description of TM-DNP. In fact both the Borghini \cite{BorghiniPRL} and the finite contact model introduced in \cite{TM1} and \cite{TM2} predict no effect of the radical concentration as long as $T_{1n}$ is negligible whereas, in presence of leakage, a higher number of polarizing centres would push the system towards higher steady state polarizations. Gadolinium doping on the other hand leaves substantially unaffected the nuclear polarization time, while shortening the typical electron spin lattice relaxation time $T_{1e}$. In the Borghini's framework such a reduction is expected to induce an enhancement of $P_n^{\infty}$, that however, as already remarked in \cite{JHAL2008}, cannot justify quantitatively those sizeable enhancements reported in the same paper and later on in \cite{JHAL2010, Lumata2012}.
To complete the experimental picture we measured the dependence of $P_n^{\infty}$ on nuclear concentration by varying the labelling percentage on a [1-$^{13}$C]pyruvic acid sample added with trityl radicals 15 mM. Even if at low nuclear concentration the polarization time becomes very long, the final value of $P_n$ remains almost constant.
\begin{figure*}[htbp]
 \includegraphics[width=17.6 cm]{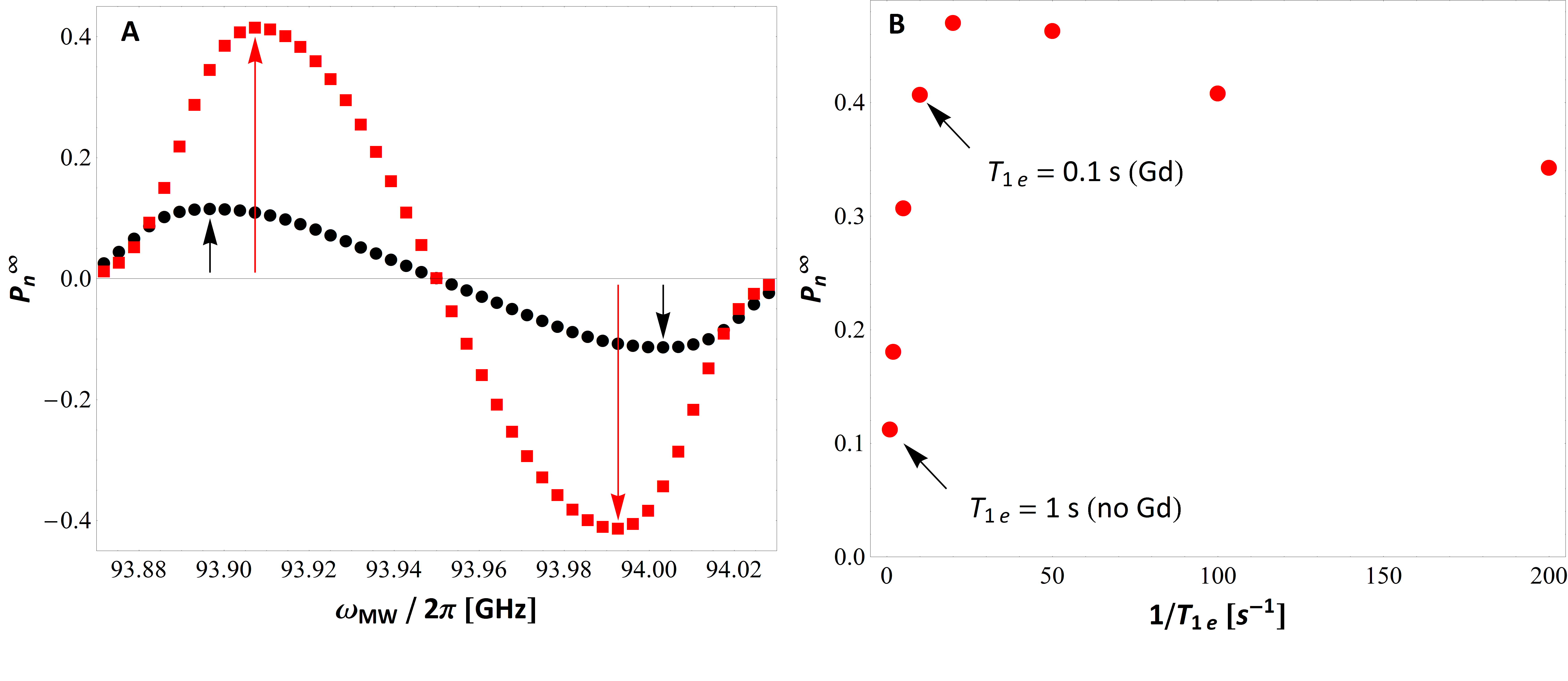}
\caption{Effect of gadolinium doping. Simulated data for a system characterized by the following parameters: $T_{\text{CSD}} = 10^{-7}$ s, $T_{\text{DSD}} = 5~10^{-5}$ s, and reproducing the behaviour of a [1-$^{13}$C]sodium pyruvate samples in a 1:1 glycerol/water glassing matrix doped with trityl 15 mM \cite{Lumata2012}. Panel A: Microwave DNP spectra upon setting $T_{1e} = 1$ s (corresponding to a sample without gadolinium \cite{JHAL2008}, circles) and $T_{1e} = 0.1$ s (corresponding to a sample with $1.5$ mM of gadolinium \cite{JHAL2008}, squares). The up and down arrows indicate the approximate positions of maximum (positive and negative respectively) polarization. Panel B. Maximum $P_n^{\infty}$ as a function of the electron relaxation rate $T_{1e}$, ranging from 5 ms to 1 s.} %(corresponding to different gadolinium concentration).}
\label{figureGdsimul}
\end{figure*}
The main messages emerging from this collection of literature and experimental observations can be summarized as follows.
\begin{itemize}
\item The nuclear spin system does not affect the evolution of the electron profile, acting only as a passive viewer of the electron system. 
\item The transition rate of many particle processes (involving nuclear and electron spins) increases with the radical concentration.
\item The addition of gadolinium complexes to the DNP preparation leads to a reduction of electron spin-lattice relaxation and to an enhancement of the steady state nuclear polarization.
\end{itemize}

Guided by these three items we introduced a novel model, based on the same rate equations approach proposed in \cite{TM1} under the assumption of bad electron-nucleus contact and negligible nuclear leakage, but including an additional mechanism accounting for energy exchanges between couples of electron spins (that flip simultaneusly) and the lattice. It is worth to mention that similar flip-flop processes were already introduced by Farrar {\em et al.} \cite{GRIFFIN} in a high temperature approximation of TM-DNP in presence of inhomogeneusly broadened ESR lines. Here we implement them within a theoretical picture where conservative spectral diffusion terms are also active and no linearization of the electron and nuclear polarization has been applied. When both the conservative and non-conservative flip-flop rates $1/T_{\text{CSD}}$  and $1/T_{\text{DSD}}$ are phenomenologically assumed to scale with $c^2$, our model is able to reproduce fairly well the experimental behaviour of $P_n^{\infty}$ on increasing the density of paramagnetic centers (see \figurename~\ref{figureApsimul} {\em versus} \figurename~\ref{figureAp}).
A quantum mechanical derivation of the two rates, and in turn of their $c$-dependence, would be desirable and certainly deserve additional future work to provide a more rigorous treatment. However, it has been checked that even if other (positive) $c$-dependences of the flip-flop rates are assumed, the qualitative outcome of this analysis remains unchanged, although the agreement between experimental and simulated data get worse. 

The implementation of the DSD, makes the model also much more sensitive to $1/T_{1e}$ variations, providing a suitable key of interpretation for the influence of gadolinium doping on nuclear polarization enhancement (see \figurename~\ref{figureGdsimul} {\em versus} \figurename~\ref{figureGd}). Remarkably the role of dissipative processes on the electron magnetic behaviour of gadolinim-doped DNP samples was anticipated by Lumata et al. \cite{Lumata2012}. Here we went over some of the assumptions limiting the theoretical analysis of the cited authors. In particular we avoided the use of the Provotorov approximation, which displays an unphysical overestimation of $P_n$ at the edges of the MW spectrum, and we extended our investigation beyond the the regime dominated by dissipative processes, where the final nuclear polarization depends monotonically on gadolinium concentration, definitely resulting in a more extensive description of the experimental scenario.

In conclusion we have presented a theoretical model providing a broad and unified understanding of different features observed in low temperature $^{13}$C DNP experiments based on the use of trityls (or narrow linewidth radicals as polarizing agents).
The validity of the model is limited to those cases where the radical concentration is high enough for TM to dominate over other polarization mechanisms such as the Solid Effect, and to make the polarization build up time fast with respect to nuclear leakage processes. 
Moreover the model is not expected to suitably describe systems characterized by broad ESR lines, where also protons are involved in the TM mechanism (as in nitroxides). In that case (that will object of future studies) the electron-nucleus contact and the nuclear leakage are likely to be strong enough to be involved in the evolution of the electron profile. 

\section{Acknowledgement}
We gratefully acknowledge L. Lumata, M.E. Merritt, C.R. Malloy, A.D. Sherry and Z. Kovacs for kindly letting us reproduce their experimental data, A. De Luca for the fruitful discussion, M. Moscardini for his technical support and Albeda Research for their contributions to DNP experiments and for the stimulating exchange of views.
This study has been supported in part by Regione Piemonte (Misura II.3 del Piano Straordinario per l'Occupazione), by the COST Action TD1103 (European Network for Hyperpolarization Physics and Methodology in NMR and MRI) and by the ANR grant 09-BLAN-0097-02.

\appendix

\section{Review of literature experimental data}
\label{review}

\begin{figure*}[htbp]
 \includegraphics[width=17.6 cm]{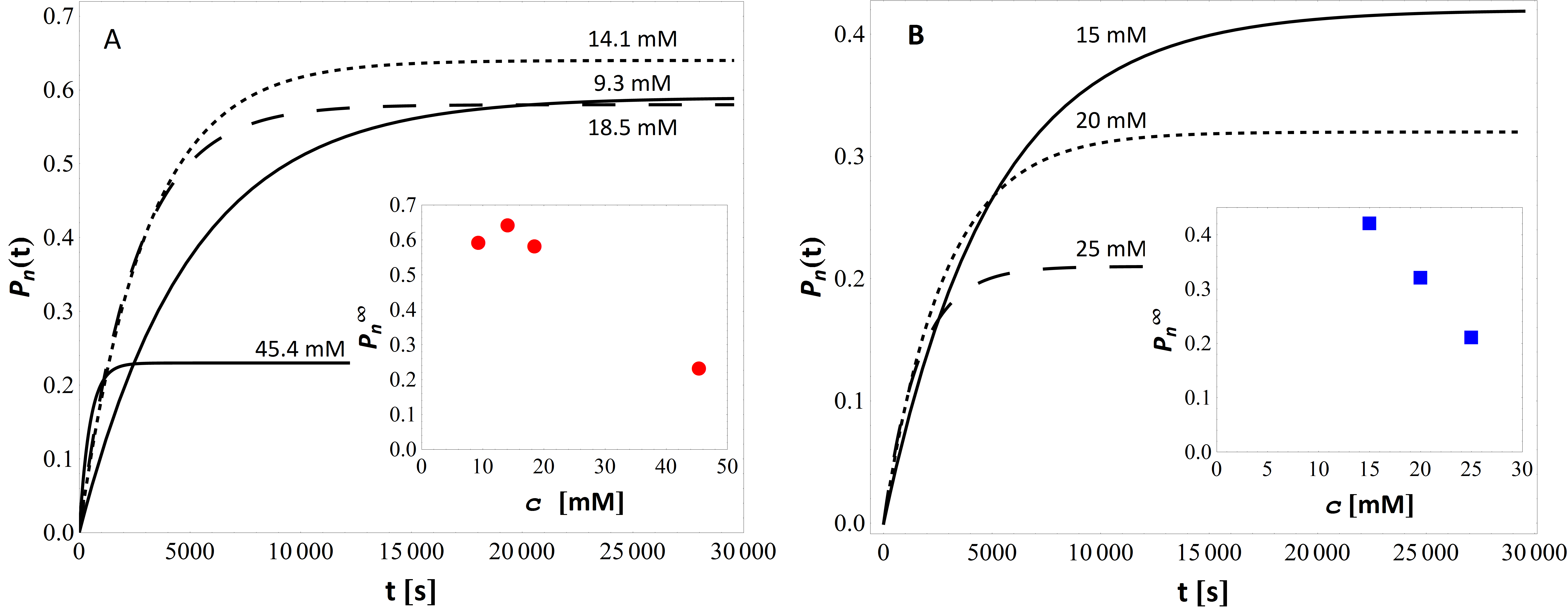}
\caption{A: Polarization build up curves and (inset) final polarization $P_n^{\infty}$ {\em versus} electron concentration c, reproduced according to the experimental data reported in \cite{JHALhighfield} acquired on [1-$^{13}$C]pyruvic acid samples at magnetic field $B_0 = 4.64$ T (corresponding to a microwave frequency of about $130$ GHz), $T = 1.15$ K and trityl concentration ranging from 9.3 to 45.4 mM. B: Build up curves and (inset) final polarization of [$^{13}$C]urea in glycerol acquired at $B_0$ = 3.35 T ({\em i.e.} about 94 GHz electron resonance frequency) and $T$ = 1.2 K with trityl concentration from 15 to 25 mM (data from \cite{Wolber2004}).}
\label{figureAp}
\end{figure*}

Different $P_n^{\infty}$ values have been observed depending on the chemical-physical properties of the investigated sample and on the type of radical used as polarizing agent.
Two compounds in particular have deserved extensive experimental studies:

\begin{itemize}
    \item $\left[\right.$1-$^{13}$C$\left.\right]$pyruvic acid, a liquid at room temperature that vitrifies upon freezing;
    \item $\left[\right.$$^{13}$C$\left.\right]$urea, a solid system at room temperature added with specific agents to promote glass formation once suddenly plunged in liquid helium.
\end{itemize}

[1-$^{13}$C]pyruvic acid has been investigated in great detail and for different experimental conditions  by Ardenkjaer-Larsen and collaborators. In particular, the behaviour of the main parameters describing the DNP process have been determined for different trityl radical concentrations \cite{JHALhighfield}. Results in terms of build up curves and polarization levels are recalled in \figurename~\ref{figureAp}, panel A. $P_n^{\infty}$ reaches its maximum value ($P_n^{\infty}$ = $0.64$) when the electron concentration is $14.1$ mM and therefore it significantly decreases on moving to $18.5$ mM ($P_n^{\infty}$ = $0.58$) and $45.4$ mM ($P_n^{\infty}$ = $0.23$). In parallel, a monotonic decrease of the polarization time $T_{\text{pol}}$ from a maximum of $5000$ s to a minimum of $475$ s is observed. A similar trend was reported in \cite{Wolber2004} for a sample of [$^{13}$C]urea, actually one of the first endogenous molecules studied for biomedical applications, dissolved in glycerol to give a nearly saturated solution. On increasing the trityl concentration from $15$ to $25$ mM (\figurename~\ref{figureAp}, panel B), the build up curves speed up while the sample experiences a strong reduction of $P_n^{\infty}$ (about a factor $2$).
Despite the different nature of the samples, a higher number of electrons is always associated to shorter $T_{\text{pol}}$ and lower $P_n^{\infty}$. 

\begin{figure*}[t]
 \includegraphics[width=17.6 cm]{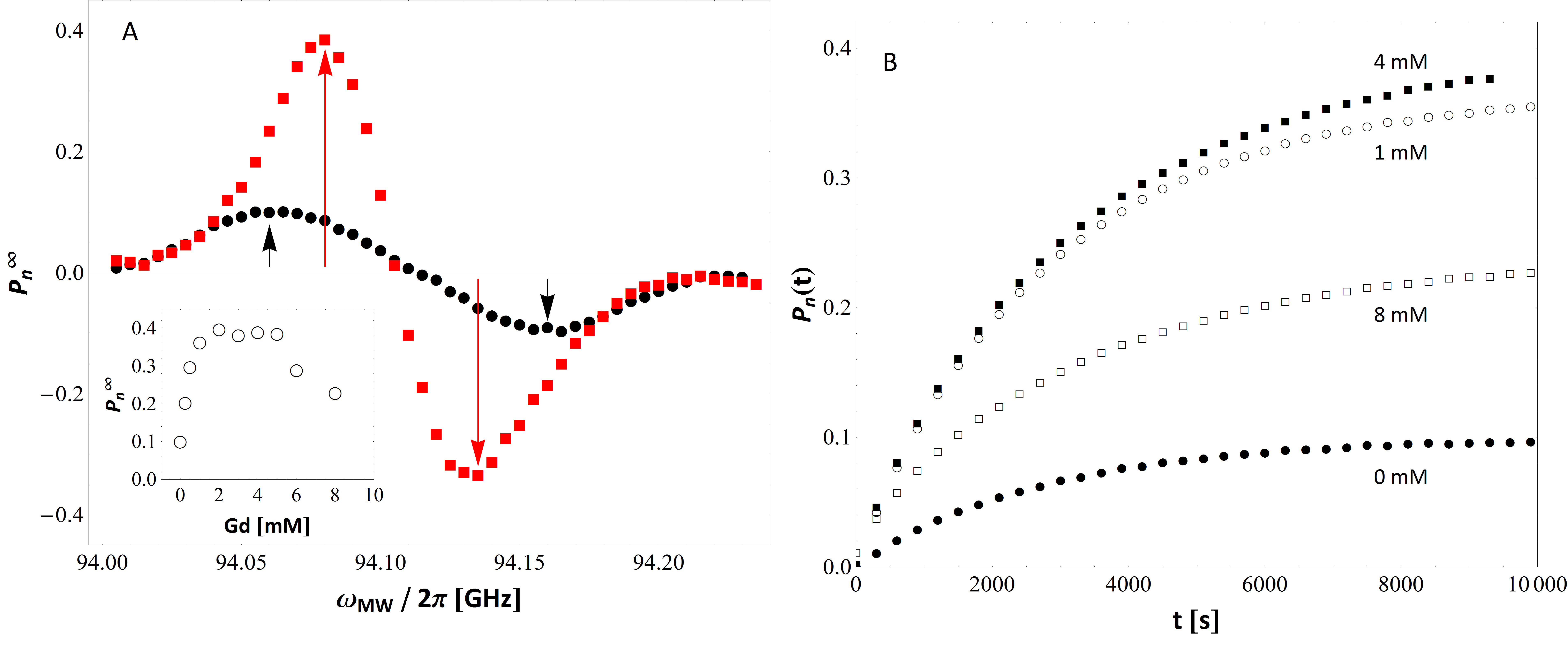}
\caption{Effect of gadolinium doping on [$^{13}$C]sodium pyruvate, data from \cite{Lumata2012}. A: Microwave spectrum for the undoped sample (circles) and for the gadolinium-$5$ mM sample (squares). The up and down arrows indicate the position of the positive and negative polarization peaks, respectively. In the inset the maximum positive polarization $P_n^{\infty}$ as a function of gadolinium concentration is represented. B: Representative polarization build up curves for different concentrations of gadolinium. The time course of polarization looks not significantly affected by the gadolinium addition. Data were collected at $3.35$ T and $1.4$ K using a $100$ mW microwave source operating at about $94$ GHz.}
\label{figureGd}
\end{figure*}

As far as the influence of the addition of small amounts of Gd-complexes on the DNP behaviour is concerned, a broad systematc study has been reported by Lumata and collaborators \cite{Lumata2012}. They investigated the effect of Gd-HP-DO3A (ProHance, Bracco Imaging) on $1.4$ M  [1-$^{13}$C]sodium pyruvate samples in a 1:1 glycerol/water glassing matrix doped with three different radical types (trityl, nitroxide, BDPA). In \figurename~\ref{figureGd} the data obtained for the solution containing $15$ mM of trityl radical and different gadolinium amounts ($0$-$8$ mM) are summarized. A comparison between the microwave spectra with and without gadolinium is shown in panel A. The addition of gadolinium leads to two main outcomes, a reduced separation between the positive and the negative polarization peaks and a nearly linear increase of $P_n^{\infty}$ as the concentration of the rare earth [Gd] goes from $0$ to $2$ mM. By further increasing [Gd], the steady state polarization first reaches a plateau and then declines for [Gd] $> 5$ mM.
The maximum $P_n^{\infty}$ achieved upon gadolinium doping is approx. $4$ times higher than what obtained in the undoped sample. Polarization times, on the other hand, were found to be only slightly affected by the gadolinium doping (\figurename~\ref{figureGd}, panel B). Similar results hold for BDPA radicals, whereas much less pronounced effects were observed when nitroxides are used for polarizing the sample. It is worth to notice that both carbon and hydrogen nuclei are involved in TM-DNP of samples prepared with nitroxides; this introduces a further degree of freedom in the problem, pushing it outside the field of validity of the theoretical model presented in the Section \ref{TheoreticalModel}.

\section{Rate equation system} \label{appA}

In order to simulate the time evolution of the electron polarization profile, the same approach proposed in \cite{TM1, TM2} has been used. The ESR line was modeled by a Gaussian function truncated at $3~\sigma$, with $\sigma = 2 \pi~27$ MHz to reproduce the behaviour of trityl based DNP samples \cite{JHAL2008}, and split into $N_p = 45$ electron packets of width $\delta \omega \approx 2 \pi~3.6$ MHz, characterized by a polarization $P_{e,i}(t)$ (indicated here with $P_{e,i}$), with $i = 1, N_p$. We have checked that our results do not change on increasing the number of packets and thus can be considered as a fair approximation of the continuum limit. The MW irradiation was assumed to be strong enough to saturate a given packet $i_0$, voiding its polarization. Then, under the assumption of weak coupling between the electron and nuclear spin systems (see the third assumption at Section \ref{TheoreticalModel}), the time evolution of the remaining packets is described by the following system of rate equations:

\begin{eqnarray}
\label{rateeqT2e}
\frac{d P_{e, i}}{dt} &=& \frac{P_0-P_{e, i}}{T_{1e}}  + \frac{1}{4 T_{\text{CSD}}} \left(f_{i-1}f_{i}f_{i+1} \Pi_{e,i} \right. \nn \\
&-& \left. \frac{1}{2} f_{i+2}f_{i+1}^2 \Pi_{e,i+1} -\frac{1}{2} f_{i-2}f_{i-1}^2 \Pi_{e,i-1} \right) \\
&+& \frac{f_{i-1}(P_{e, i-1}-P_{e, i})+f_{i+1}(P_{e, i+1}-P_{e, i})}{T_{\text{DSD}}}  \nn 
\end{eqnarray}
where $P_0 = \tanh \left[ \beta_L \omega_e\right]$ ($\omega_e = \gamma_e B_0$, with $\gamma_e = - 2 \pi~28.025$ GHz/T) and $\Pi_{e,i}$ is given by the expression:
\begin{equation}
\Pi_{e,i} = \left(P_{e, i-1}\small{+}P_{e, i+1} \right)\left(1\small{+}P^2_{e, i}\right) - 2 \left(1\small{+}P_{e, i-1}P_{e, i+1}\right)P_{e, i}.\nonumber 
\end{equation}

The multi-particle evolution terms in \ref{rateeqT2e} are formally derived in mean field approximation in Appendix \ref{appC}. 
The system \ref{rateeqT2e} has been numerically solved to compute $P_{e,i}$ imposing a discrete time step $dt$, so that:

\begin{equation}
\frac{1}{dt} = \frac{1}{T_{1e}}+\frac{\sum_i f_i^2f_{i-1}f_{i+1}}{2 T_{\text{CSD}}}+ \frac{\sum_i f_i f_{i+1}}{T_{\text{DSD}}}. \nn
\end{equation}

The function $P_e(\omega, t)$ to be used in Eq.(\ref{eqQ}) was thus obtained by imposing $\omega = \omega_0 + (i- \frac{N_p + 1}{2})~\delta \omega$ and $\omega_n = 9~\delta \omega \approx 2 \pi~35$ MHz which corresponds to the $^{13}$C Larmor frequency at $B_0 =  3.35$ T.
Finally, the steady state polarization $P_e^{\infty}(\omega)$ for Eq. \ref{eqpn*} has been calculated as the limit of  $P_e(\omega, t)$, for $t \rightarrow \infty$.

\section{Mean field derivation of $T_{\text{CSD}}$ and $T_{\text{DSD}}$ evolution terms} \label{appC}

\emph{Conservative spectral diffusion}

In the mean field approximation, the number of possible conservative spectral diffusion processes involving four electrons is given by $\sum_i \frac{1}{2} N_e^4 f_i^2 f_{i-1}f_{i+1}$. The total rate needs to linearly scale with the volume of the system in order to assure a correct thermodynamical limit. To achieve this, as usually done for fully connected models, the effective time constant of each four particle process must depend on the system size and scale as:
\begin{equation}
T_{\text{CSD}}^{\text{eff}}=T_{\text{CSD}}N_e^3,
\end{equation}
where the constant $T_{\text{CSD}}$ is size indipendent. Then the total rate of all spectral diffusion events writes $W_{\text{CSD}} = N_e \sum_i \frac{1}{2} f_i^2 f_{i-1} f_{i+1} / T_{\text{CSD}}$.

\begin{figure}[htbp]
 \includegraphics[width=8cm]{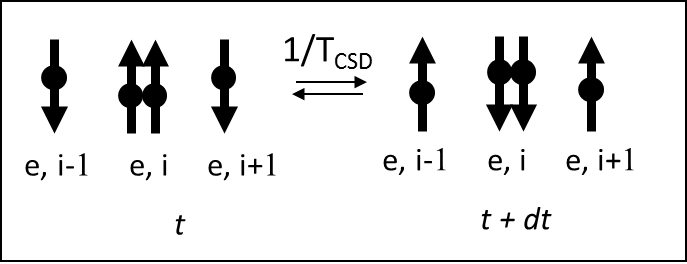}
\caption{Schematic representation of a conservative spectral diffusion event involving two spin flips in packet {\em i}.}
\label{figureT2e}
\end{figure}

Let us now define $P_{e, i}^+$ as the fraction of electrons $up$ belonging to the packet $i$ and $P_{e,i}^-$ the fraction of electrons $down$ belonging to the packet $i$.
When the conservative spectral diffusion event depicted in \figurename~\ref{figureT2e} (where two spins of packet {\em i} flip) occurs, the fraction of electrons $up$ in the $i$-th packet is decreased by $2/(N_e f_i)$.
The number of possible transitions is the product of:
\begin{itemize}
	\item the number of the electrons $down$ in the $(i-1)$-th packet: $N_e f_{i-1} P_{e,i-1}^-$,
	\item the number of the electron pairs $up$ in the $i$-th packet: $\frac{1}{2} N_e^2 f_{i}^2 (P_{e, i}^+)^2$,
	\item the number of the electrons $down$ in the $(i+1)$-th packet: $N_e f_{i+1} P_{e,i+1}^-$. 
\end{itemize}
The rate of such process is $1/ (T_{\text{CSD}} N_e^3)$, and the total decrement of $P_{e, i}^+$ in the time interval $dt$ is:
\begin{equation} 
-\frac{dt}{T_{\text{CSD}}} f_{i-1}f_{i}f_{i+1} P_{e,i-1}^- (P_{e, i}^+)^2 P_{e,i+1}^-. 
\end{equation}
The variation of $P_{e, i}^+$ induced by all possible electron spectral diffusion transitions, similar to the one showed in \figurename~\ref{figureT2e}, is given by:
\begin{equation} 
\frac{dt}{T_{\text{CSD}}} f_{i-1}f_{i}f_{i+1}\left[\small{-}P_{e,i-1}^- (P_{e, i}^+)^2 P_{e,i+1}^-\small{+}P_{e,i-1}^+ (P_{e, i}^-)^2 P_{e,i+1}^+ \right]
\end{equation}
\begin{figure}[htbp]
 \includegraphics[width=8cm]{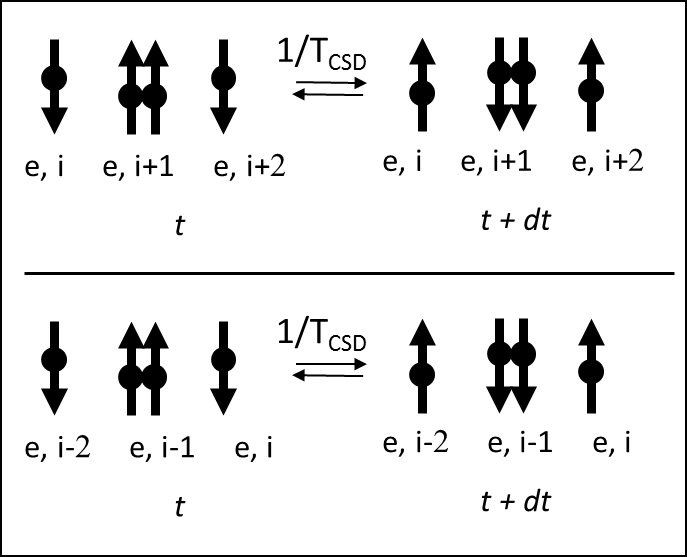}
\caption{Schematic representation of conservative spectral diffusion events involving one electron flip in packet {\em i}.}
\label{figureT2ebis}
\end{figure} 
Two other different transitions have to be considered, involving a single spin flip in packet {\em i} (\figurename~\ref{figureT2ebis}). 
By using the same approach described above, the total variation of $P_{e, i}^+$ will also contain the two following terms:
\begin{eqnarray} 
\frac{dt}{2T_{\text{CSD}}} f_{i+1}^2f_{i+2}\left[ P_{e,i}^- (P_{e, i+1}^+)^2 P_{e,i+2}^- - P_{e,i}^+ (P_{e, i+1}^-)^2 P_{e,i+2}^+ \right]\nonumber \\
\frac{dt}{2T_{\text{CSD}}} f_{i-1}^2f_{i-2}\left[ P_{e,i-2}^- (P_{e, i-1}^+)^2 P_{e,i}^- - P_{e,i-2}^+ (P_{e, i-1}^-)^2 P_{e,i}^+ \right]\nonumber
\end{eqnarray}
Using the relations: 
\begin{eqnarray}
\label{ppp}
P_{e, i}^+=\frac{(1+P_{e, i})}{2}, \nonumber \\
P_{e, i}^-=\frac{(1-P_{e, i})}{2},  
\end{eqnarray}
the total variation of $P_{e, i}$ induced by all possible spectral diffusion processes, $\delta P_{e, i} =2 \delta P_{e, i}^+$, can be easily derived by means of simple algebraic calculations.

\emph{Dissipative spectral diffusion}

The number of the possible flip-flops involving two electrons is given by $\sum_i f_i N_e f_{i+1} N_e$. 
To properly assure the correct thermodynamic limit, the effective time constant of each two particle process must depend on the system size and scale as:
\begin{equation}
T_{\text{DSD}}^{\text{eff}}=T_{\text{DSD}} N_e,
\end{equation}
where the constant $T_{\text{DSD}}$ is size indipendent. 

Then the total rate of all $T_{\text{DSD}}$ events writes $W_{\text{DSD}} = N_e \sum_i f_i f_{i+1} / T_{\text{DSD}}$.

\begin{figure}[htbp]
 \includegraphics[width=7cm]{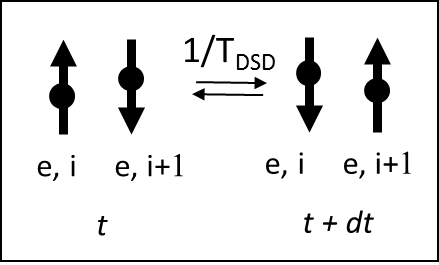}
\caption{Schematic representation of one possible dissipative spectral diffusion event.}
\label{figureTF}
\end{figure}

When the dissipative spectral diffusion event depicted in \figurename~\ref{figureTF} occurs, the fraction of electrons $up$ in the $i$-th packet is increased by $1/(N_e f_i)$.
The number of possible transitions is the product of:
\begin{itemize}
	\item the number of the electrons $down$ in the $i$-th packet: $N_e f_i P_{e,i}^-$,
	\item the number of the electrons $up$ in the $(i + 1)$-th packet: $N_e f_{i+1} P_{e, i + 1}^+$. 
\end{itemize}
The rate of such process is $1/ (T_{\text{DSD}} N_e)$, and the total increment of $P_{e, i}^+$ in the time interval $dt$ is:
\begin{equation} 
\frac{dt}{T_{\text{DSD}}} f_{i+1} P_{e,i}^- P_{e, i + 1}^+. 
\end{equation}
The total variation of $P_{e, i}^+$ induced by all possible electronic flip-flop transitions, $\delta P_{e, i}^+$, is given by:
\begin{eqnarray} 
\delta P_{e, i}^+ &=& \frac{dt}{T_{\text{DSD}}} \left[ f_{i+1} \left( P_{e,i}^- P_{e, i + 1}^+  - P_{e,i}^+ P_{e, i + 1}^-\right) \right. \nn \nonumber \\
&& \left. \small{+} f_{i-1} \left( P_{e,i}^- P_{e, i - 1}^+  - P_{e,i}^+ P_{e, i - 1}^- \right)\right]\nonumber. 
\end{eqnarray}
Using the relations \ref{ppp}, the total variation of $P_{e, i}$ induced by all possible flip-flop processes, $\delta P_{e, i} =2 \delta P_{e, i}^+$, can be written as follows:
\begin{eqnarray} 
\label{RE1TF}
\delta P_{e, i} &=& \frac{dt}{2 T_{\text{DSD}}} \left\{ f_{i+1} \left[ (1\small{-}P_{e,i}) (1\small{+}P_{e, i + 1})  \right. \right. \nn \nonumber \\
&& \left. - (1\small{+}P_{e,i})(1\small{-}P_{e, i + 1})\right] \nonumber \\
&& \small{+} f_{i-1} \left[ (1\small{-}P_{e,i})(1\small{+}P_{e, i - 1}) \right. \nonumber \\
&& \left. \left. -(1\small{+}P_{e,i})(1\small{-}P_{e, i - 1})\right] \right\}. 
\end{eqnarray}
The term proportional to $1/T_{\text{DSD}}$ in the system of rate equations \ref{rateeqT2e} can be now easily derived from Eq. \ref{RE1TF}.

\end{document}